\documentclass[aps,prb, twocolumn]{revtex4}

\usepackage{graphicx}
\usepackage{amssymb}
\usepackage{amsmath}
\usepackage{colordvi}
\usepackage{color}

\newcommand{\rev}[1]{\textcolor{black}{#1}}
\newcommand{\rerev}[1]{\textcolor{black}{#1}}

\begin{document}

\title{Multistable localized states in \rerev{highly photonic}  polariton rings with a  quasiperiodic modulation}

\author{Andrei V. Nikitin and Dmitry A. Zezyulin}

\affiliation{School of Physics and Engineering, ITMO University, St. Petersburg 197101, Russia}

%\affil[*]{d.zezyulin@gmail.com}

\begin{abstract}
We present a theoretical study of  an exciton-polariton annular microcavity  with an additional quasiperiodic structure along the ring which is implemented in the form of a bicosine dependence.   We demonstrate that for  a sufficiently strong quasiperiodic modulation, the microcavity features a sharp mobility edge separating a cluster of localized states from the rest of the spectrum consisting of   states extended over the whole ring. Localized modes can be excited using a resonant pump whose topological charge determines   the phase distribution of excited patterns. Repulsive polariton interactions  make the resonance peaks distinctively asymmetric and enable  the formation of multistable states which feature the attractor-like dynamical behavior \rev{and hysteresis}.  We also demonstrate that  the localized states   can be  realized in a biannular cavity that consists of two rings, each having periodic modulation, such that the periods of two modulations are different. 
\end{abstract}

\maketitle

\section{Introduction}

Cavity polaritons are composite light-matter quasiparticles that emerge in  a planar semiconductor microcavity  operating in the so-called strong-coupling regime \cite{Deng,Soln2021}.  From their photonic component, cavity polaritons have macroscopically large   coherence length, while the excitonic fraction enables strong polariton-polariton interactions. Another peculiarity of exciton-polaritons is their inherently nonequilibrium nature due to the photon leakage from the cavity. A quasistationary regime  can be achieved when   losses are   compensated  with a resonant or nonresonant pump. Exciton-polaritons can be confined and controlled by  external  potentials of various shapes. State-of-the-art nanotechnology provides a variety of   polariton trapping techniques and various means for  controllable energy   landscape   engineering  \cite{Schneider}. In particular, polaritons can be confined in    nonsimply-connected, i.e., ring- or multiring-shaped geometries  \cite{Dreismann,Snoke} which facilitate  the manifestation of  the superfluid nature of polariton condensates \cite{Carusotto} through the formation of  vortex states,   persistent circular currents, and  rotating patterns, see e.g. Refs.~\cite{LiFraser,Barkhausen2020,Barkhausen2021,Bochin,Gnusov,Redondo,Lukoshkin,Yulin,Kartashov} .   In addition, exciton-polaritons undergo condensation   at relatively high temperatures    as compared to   Bose condensates of cold atoms.  Combination of these properties makes exciton-polaritons  an ideal platform for   studies  of quantum collective phenomena in an environment  where nonlinear and nonequilibrium effects play a prominent role. 

A remarkable example of such a phenomenon is the spatial  localization  in incommensurate lattices.  While this behavior  is known for more than 40~years, at least since the pioneering work of Aubry and Andr\'e (AA) \cite{AA},   quasicrystalline systems  and a host of associated physical phenomena --- such as the existence of the mobility edge, `metal-insulator transition', and the fractal energy spectrum ---  continue to attract steadily growing attention in various areas of physics. For  cavity polaritons, the  localization in quasiperiodic structures has been  experimentally  observed   in   layered systems based on  the one-dimensional polaritonic wires structured in the form that interpolates between the AA and  Fibonacci sequences \cite{Tanese, Goblot}.  \rev{In the experiments reported on in Ref.~\cite{Goblot}, the polaritonic wires were excited with   a weak non-resonant cw laser, and spatially extended and localized states were identified using the spectrally resolved photoluminescence  signal either in real and in $k$ space.} In the meantime,   the theoretical treatment of exciton-polaritons that undergo the localization  in quasiperiodic landscapes   has mostly  been   limited by the linear and conservative limit, and a complete account of the  strong polariton nonlinearity and resonant   pump has not yet been performed.  

The main goal of the present work is to explore the effect of resonant pump and nonlinearity on the formation and stability of  localized states in   a structured microcavity.  We treat the problem using the so-called approximant path which replaces the quasiperiodic system with a long-periodic one \cite{Diener,LiLiSarma,Zilberberg}. This transformation  suggests  considering  the  system   under the periodic boundary conditions, and therefore  the annular  geometry becomes the most natural one. We model a polariton condensate   confined in an annular-shaped cavity which has an additional quasiperiodic structuring approximated using a bicosine landscape.  \rerev{It is assumed   that the condensate has  a high photonic fraction,  which makes the phonon-assisted energy relaxation less efficient}. We show that this system can feature a sharp mobility edge that separates delocalized and localized states. A peculiarity of the adopted approach   is that the number of localized states can be controlled by the chosen rational approximation (at least, in a certain  range of parameters). Another  representative feature of the annular  geometry is the possibility to excite    localized states  using a vortical ring-shaped  resonant pump which produces a number of closely spaced resonances. Vorticity of the pump can be used to control the phase portrait of localized patterns.  In the nonlinear regime, the resonant peaks become strongly asymmetric, which  results in     multistable localized states coexisting at the same  the pump frequency.  Dynamical simulations reveal that each   unstable state dynamically switches to  its   stable counterpart situated in the   upper branch  of the multistable resonance curve. The existence of localized states is illustrated for two configurations: the first one corresponds to a single ring with an additional biperiodic modulation, and the second one is a biannular structure, where the two rings have modulations with different periods.

The organization of the paper is as follows. In Sec.~\ref{sec:model} we introduce the adopted theoretical model. In Sec.~\ref{sec:linear} we describe localized states in the linear and conservative regime (i.e., when the resonant pump and losses are neglected). In Sec.~\ref{sec:resonances} we present the  results on the resonant excitation of linear and nonlinear states and demonstrate that in the latter case the multistability takes place.  Section~\ref{sec:resonances} is devoted to the excitation of out-of-phase localized states using a resonant pump with  a properly chosen vorticity. Section~\ref{sec:concl} provides a conclusion and a short outlook.

\section{Model}
\label{sec:model}
A well-known possibility to realize  the   quasiperiodicity corresponds to a bichromatic potential comprising two  periodic potentials of incommensurate periods.  While the AA model implements such a situation in a tight-binding approximation, the simplest   example of a spatially-continuous  one-dimensional  quasiperiodic potential  is  a bicosine landscape $V_0[ \cos z + \cos (\varphi z)]$, where $\varphi$ is an irrational number,   $z \in \mathbb{R}$ is a spatial coordinate, and $V_0$ is the amplitude. Complete understanding of even such a  relatively simple potential can be an overwhelmingly complex task, see e.g. \cite{Frolich,Simo,Cohen,Diener,Boers,Modungo,Biddle,LiLiSarma,Yao}.  One of possible approaches relies on the  use of   a rational approximation $\varphi \approx M/N$, where $M \gg1 $ and $N \gg1$ are coprime integers. This approach proceeds by    replacing the quasiperiodic potential by the following one: 
\begin{equation}
\label{eq:1D}
V_{\mathrm{1D}}(z) = V_0[\cos z + \cos(M z/N)].
\end{equation}
The latter potential is \emph{periodic} with the period equal to $2\pi N$, and therefore it cannot   support truly localized states, i.e., eigenstates that satisfy the zero boundary conditions at the infinity. However 
%, because its spectrum consists of   Bloch waves.
the concept of localization can be nevertheless introduced  by considering   Bloch waves that are strongly  localized within the unit cell   $[-\pi N, \pi N)$. Such an \textit{approximant path} \cite{Zilberberg}  allows one to reach the physics of quasicrystals using  the well-developed   Fourier-Floquet-Bloch machinery and, moreover, enables   the concept of topological quasicrystals \cite{Zilberberg}. In particular, topological invariants of the resulting minibands (such as Chern  numbers or Zak phases) can be connected to the spatial distribution of  the localized states \cite{ZezKon2022}.  

The   transformation from a quasiperiodic system to a large-periodic one suggests  considering  the latter one under the periodic boundary conditions. Therefore, the annular  geometry becomes the most natural one for  such a system.  \rerev{In this study, we consider a system of exciton-polaritons formed by cavity photons coupled to quantum well excitons. We assume that polaritons are confined in a structured ring-shaped cavity and their dynamics can be described by a single classical field for lower-branch polaritons.  Then the dynamics can be modelled  using  the coherently driven   Gross-Pitaevskii equation }\rev{\cite{Carusotto}} (GPE) 
\begin{equation}
\label{eq:gpe}
i\psi_t =  \left[ -\frac{1}{2}(\partial_x^2 + \partial_y^2)     + V  + |\psi|^2  - i\gamma \right] \psi  + h(r)e^{im\theta -i\varepsilon t}.
\end{equation}

Equation  (\ref{eq:gpe})  is written in normalized units: for the effective polariton mass $m\approx 10^{-34}$~kg and the unit length $\ell \approx 1~\mu$m, the adopted normalization corresponds to the time unit $m\ell^2/\hbar\approx 1$~ps and   the energy unit $\hbar/(m\ell^2)\approx 0.7$~meV. Function $h(r)$ describes the spatial modulation of resonant pump (hereafter $r\geq 0$ and $\theta\in [-\pi, \pi)$ are the polar coordinates). We assume that $h(r)$ is characterized by the maximal amplitude $h_{0}$ and has annular shape with the same radius $r_0$ as that of the ring-shaped microcavity, and the width of the annular pump is much  larger than that of the cavity ring,  i.e.,  the  intensity of the pump beam is almost uniform over the cavity.  Integer $m$ is the topological charge of the pump beam.  The nonlinear term $\propto |\psi|^2$ corresponds to polariton interactions. Small coefficient $\gamma$ accounts for polariton losses (we use model value $\gamma\approx 0.02$).
%Coefficient $g\geq 0$ governs the strength of the repulsive interactions. Obviously one of the parameters (either $g$ or $h_{max}$) can be considered as a fixed one.
%\begin{equation}
%\label{eq:GPE}
%i\psi_t = \left[ -\frac{1}{2}\nabla^2   + V  + g |\psi|^2  - %i\gamma\right] \psi  + H(x,y, t).
%\end{equation}
Function $V$ describes the landscape of the confining potential. To  specify its shape, we for a moment return   to    the $2\pi N$  one-dimensional periodic potential (\ref{eq:1D})    and    map  its unit cell  $[-\pi N, \pi N)$ to the period   of the polar angle, i.e., to  $[-\pi, \pi)$. A simple transformation $\theta=z/N$   leads to the following bicosine dependence: $V_0[\cos (N\theta ) + \cos (M\theta)]$. We therefore use   the following  landscape for potential $V$ in Eq.~(\ref{eq:gpe}): 
\begin{equation}
\label{eq:V}
V(r, \theta) = V_R(r)[1 + \delta(\cos(N\theta) + \cos(M\theta)) ].
\end{equation}
Here function $V_R(r)$ models the uniform ring potential: $V_R(r) = -V_0  e^{-(r-r_0)^2/w_0^2}$,
%V_R(r) = -V_0 [e^{-(r-r_0)^2/w_0^2} + e^{-(r+r_0)^2/w_0^2} ]$, 
where $V_0$, $r_0$, and $w_0$ are positive parameters that tune depth of the annular potential, its radius, and width, respectively. Coefficient $\delta\in(0,1)$ is  the strength of the quasiperiodic modulation  in units of the average  depth $V_0$. In recent experiments, the energy landscape gradient along the ring has been realized by the variation of the thickness of the microcavity \cite{Mukherjee}. Several  proposals on the implementation of a    modulated annular cavities have been recently outlined in Ref.~\cite{Chestnov}.  In particular, a circularly distributed modulation can be realized with a  structured optical mask imaged onto the sample \cite{Dall,Wurdak}.

\begin{figure*}[ht]
\centering
{\includegraphics[width=\linewidth]{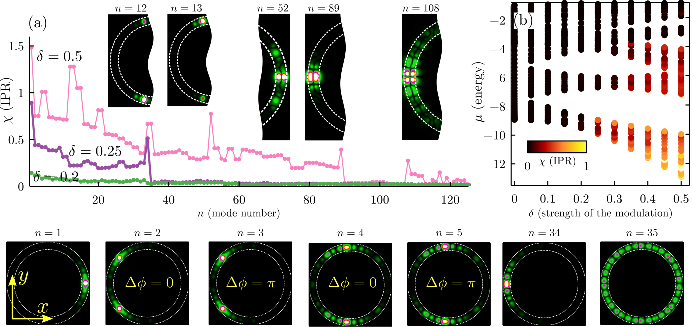}}
\caption{(a) IPR $\chi$ for 125 lowest linear eigenstates (enumerated with index $n=1,2,\ldots$) for three different strengths of the quasiperiodic modulation: $\delta=0.2, 0.25$, and $0.5$. (b)  Energy $E$ vs.     $\delta$ for lowest eigenvalues. Different colors correspond to different values of the IPR. Seven lowest panels: moduli distribution $|\psi(x,y)|$ for several representative eigenstates for $\delta=0.25$. Labels $\Delta\phi=0$ and $\Delta\phi=\pi$ correspond to in-phase and out-of-phase double-peaked states, respectively. Upper diagrams: several moduli distributions for $\delta=0.5$: a pair of asymmetric localized states ($n=12,13$), two double states ($n=52, 89)$, and a quadruple state ($n=108$);  for compactness of the figure, only part of the ring is shown for these states. In this picture radius $r_0 = 10$, squared meanwidth $w_0^2 = 0.5$, average depth $V_0=12$; $M = 55$, $N=34$.}
\label{fig:linspectrum}
\end{figure*}

\rev{We note that the coherently driven GPE equation (\ref{eq:gpe}) is a simplified model which is generally valid only if the Rabi frequency is large as compared to other relevant energy scales \cite{Carusotto}. In addition, Eq.~(\ref{eq:gpe}) disregards  the polariton energy relaxation   that  is   known to be prominent in certain polaritonic systems \cite{Wouters1,Wouters2,Savenko,Winkler}. Nevertheless this model still can be useful to predict the very existence of localized states.  The description provided by Eq.~(\ref{eq:gpe}) is expected to be more accurate   for highly photonic condensates where the phonon-assisted energy relaxation becomes less  efficient.}
\rerev{An equation of mathematically similar structure (known as the Lugiato-Lefever equation \cite{Lugiato}) is used to model light dynamics in planar cavities filled with a nonlinear Kerr medium. Recent experiments (see e.g. \cite{Lucas,Moille}) report on the possibility of patterning a  ring-shaped optical microresonator with implementing a photonic crystal or  a multifrequency photonic crystal  along   the microring. The results of our study can   be relevant in this subfield as well.   }

\section{Linear localized states in the conservative regime}
\label{sec:linear}

\begin{figure}
	\begin{center}
		\includegraphics[width=1\columnwidth]{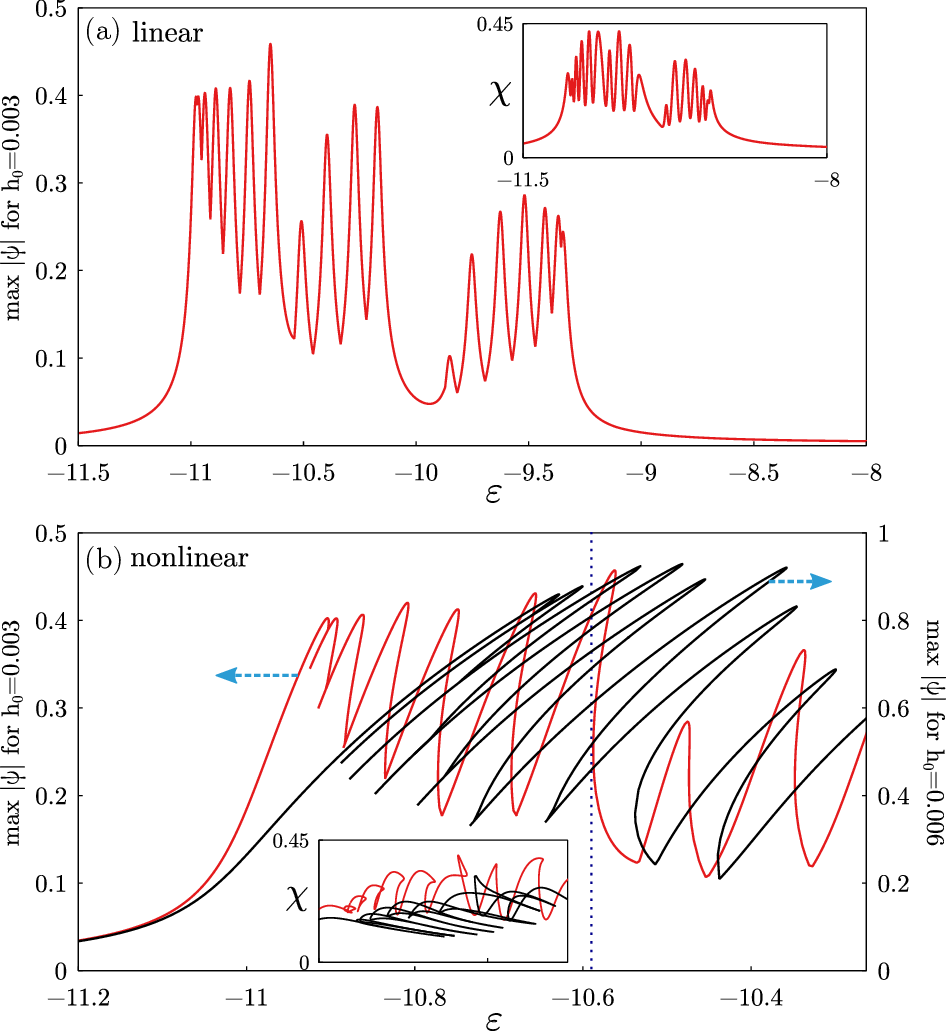}%
	\end{center}
	\caption{(a):  Maximal amplitude $\max_{(x,y)} |\psi|$ for linear states excited by the resonant pump with maximal amplitude  $h_0 = 0.003$. (b):  Resonances in the nonlinear regime for two values of  $h_0$ ($h_0=0.003$ and $h_0=0.006$ corresponding to the left and to the right vertical axis, respectively); notice that in this panel only part of all resonances is displayed.  The insets show IPRs $\chi$. In this figure, the pump is   zero-charged, i.e.,  $m=0$.}
	\label{fig:linresonances}
\end{figure}

The spectrum of stationary eigenstates $\psi =  e^{-i \mu t} u(x,y)$ has been found numerically  from the two-dimensional eigenvalue problem obtained from Eq.~(\ref{eq:gpe}) in the regime  where nonlinearity, losses $\gamma$, and pump $h(r)$  are  switched off.   To model the potential, we have used   $M=55$ and $N=34$ in Eq.~(\ref{eq:V}).  These   two subsequent Fibonacci numbers correspond to a rational approximation of the golden  ratio $M/N \approx \varphi=(1+\sqrt{5})/2$.  To quantify the localization of eigenmodes, we use the inverse participation ratio (IPR) defined as (see e.g. \cite{Goblot,Yao}) 
\begin{equation}
\chi =   \frac{\iint_{-\infty}^\infty |\psi|^4dxdy} {( {\iint_{-\infty}^\infty  |\psi|^2dxdy})^2}.
\end{equation}
In Fig.~\ref{fig:linspectrum}(a) we present IPRs computed for 125 lowest-energy  states for three different values of the modulation strength $\delta$. For $\delta\lesssim 0.2$ all states are delocalized or only weakly localized, i.e., extended along the ring (with IPRs below 0.15). In this case  a distinctive mobility edge, i.e., a sharp boundary between localized and delocalized states, cannot be found in the spectrum. However, at $\delta=0.25$ we observe that there are exactly 34 localized states (with IPRs $\gtrsim  0.2$)  which are situated at the bottom of the spectrum and  are   sharply distinguished from the delocalized modes (compare panels $n=1, 2, \ldots 34$ with $n=35$ in the lowest row of Fig.~\ref{fig:linspectrum}). Thus, similar to the earlier studies  where the rational approximation was used for one-dimensional quasiperiodic potentials  (see in particular \cite{ZezKon2022, Prates}), the exact number of localized states can be controlled by  the choice of the adopted rational approximation.  In the one-dimensional quasiperiodic potential (\ref{eq:1D}) this behavior can be explained by the fact that   each spectral band of the $2\pi$-periodic potential associated with the first cosine splits into $N$ \emph{minibands} in the spectrum   of the full  $2 \pi N$-periodic potential comprising both cosines.  From the pseudocolor diagram in Fig.~\ref{fig:linspectrum}(b)  we observe that for $\delta=0.25$ there is a relatively large gap  between the energies of localized and delocalized states (again, this agrees with the earlier results for one-dimensional lattices \cite{Yao}). A closer inspection of localized states shows that there are two single-peak states (corresponding to $n=1$ and $n=34$), where the location of the peak corresponds to $\theta=0$ and $\theta=\pi$, respectively. Due to the symmetry of the potential with respect to the angular reversal   $\theta\to-\theta$, other localized states  consist of pairs of   in-phase and out-of-phase two-peaked states having close energies,  with the phase difference between the peaks equal to $\Delta\phi=0$ and $\Delta\phi=\pi$, respectively, see panels $n=3,4$ and $n=4,5$ in the lowest row of Fig.~\ref{fig:linresonances}. For stronger modulation (such as $\delta=0.5$), the picture becomes more complex. In particular, there appear pairs of asymmetric states which are degenerate, i.e., have equal energies (notice that localized states of this type are impossible in the   strinctky one-dimensional geometry subject to the zero boundary conditions due to the well-known fact that  the one-dimensional Schr\"odinger equation cannot have   double eigenvalues associated with localized eigenstates \cite{LL}).  Moreover, a sufficiently strong modulation enables localization of more complex states that belong to upper polariton subbands and feature a distinct structure along the radial coordinate    (see e.g. double and quadruple states displayed in  panels $n=52, 89, 108$ in Fig.~\ref{fig:linspectrum}).  Similar states with complex transverse distribution have been detected experimentally in quasi-1D polariton wires \cite{Goblot}.  As a result of the excitation of the upper subbands, the strongly  modulated potential supports more localized states and can have multiple mobility edges corresponding to the \emph{reentrant} localization transition, see the interval $n=100\ldots120$ for $\delta=0.5$ in Fig.~\ref{fig:linspectrum}(a).  The first transition from localized to delocalized states corresponds to $n=89$ which is another Fibonacci number.

 %We have four types of localized states: single-peaked state with the peak situated exactly at $\theta=0$ (?), states composed of two identical inphase peaks or two   out-of-phase peaks, and double (``degenerate'') asymmetric states corresponding to the map $\theta \to -\theta$ 
 
%For each run, we save the dependence $\max |\Psi(x,t)| $ and $N_+ = \int_x$

\begin{figure}
	\begin{center}
		\includegraphics[width=\linewidth]{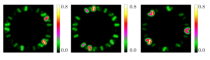}%
	\end{center}
	\caption{Examples of localised states in the biannular potential that consists of two structured rings, see Eq.~(\ref{eq:biann}). Here $N=21$, $M=34$, $\delta_M=0.45$ and $\delta_N=0.35$. Radii of the rings are $r_0=10$ and $r_0-\Delta r = 8.6$.}
	\label{fig:tworings}
\end{figure}

Concluding the discussion of localization if  the linear and conservative regime, we notice that the  excitation of localized states is also possible in    a \emph{biannular} structure, where each ring has a {periodic} modulation, and the periods are different for both rings. The corresponding potential landscape can be   represented as 
\begin{eqnarray}
\label{eq:biann}
V=V_R(r)[1+\delta_M\cos(M\theta)] \hspace{3cm}\nonumber\\[1mm]
+ V_R(r-\Delta r)[1+\delta_N\cos(N (\theta +\theta_0))],
\end{eqnarray}
where $\Delta r>0$ is the difference between the ring radii,   $\theta_0$ is the angular shift  between the two rings, and $\delta_{M,N}$ are the modulation strengths for larger ($M$) and smaller ($N$) rings.  Several examples of localized states found in such  a structure are displayed in Fig.~\ref{fig:tworings}.

\rev{\section{Resonances,  multistability, and hysteresis-like behavior}}
\label{sec:resonances}

Next, we proceed to  the localized states in the presence  of  the resonant pump and polariton nonlinearity. We focus on the cavity with the single structured ring     described by potential (\ref{eq:V}).  When the pump and losses are switched on  (in the so far linear system), the localized   states produce a picture with multiple coexisting resonances which is shown in   Fig.~\ref{fig:linresonances}(a) as a dependence of maximal amplitude  of the excited pattern on the pump frequency $\varepsilon$. The spectrum of IPRs (shown in the inset) features a similar resonant structure.   The total number of resonant peaks in   Fig.~\ref{fig:linresonances}(a) is less than the number of localized states that can be found in the conservative regime. This is due to the fact the the zero-charged resonant pump cannot excite   out-of-phase states (this issue will be discussed below in Sec.~\ref{sec:out}).

\begin{figure}
 	\begin{center} 		
    \includegraphics[width=1\linewidth]{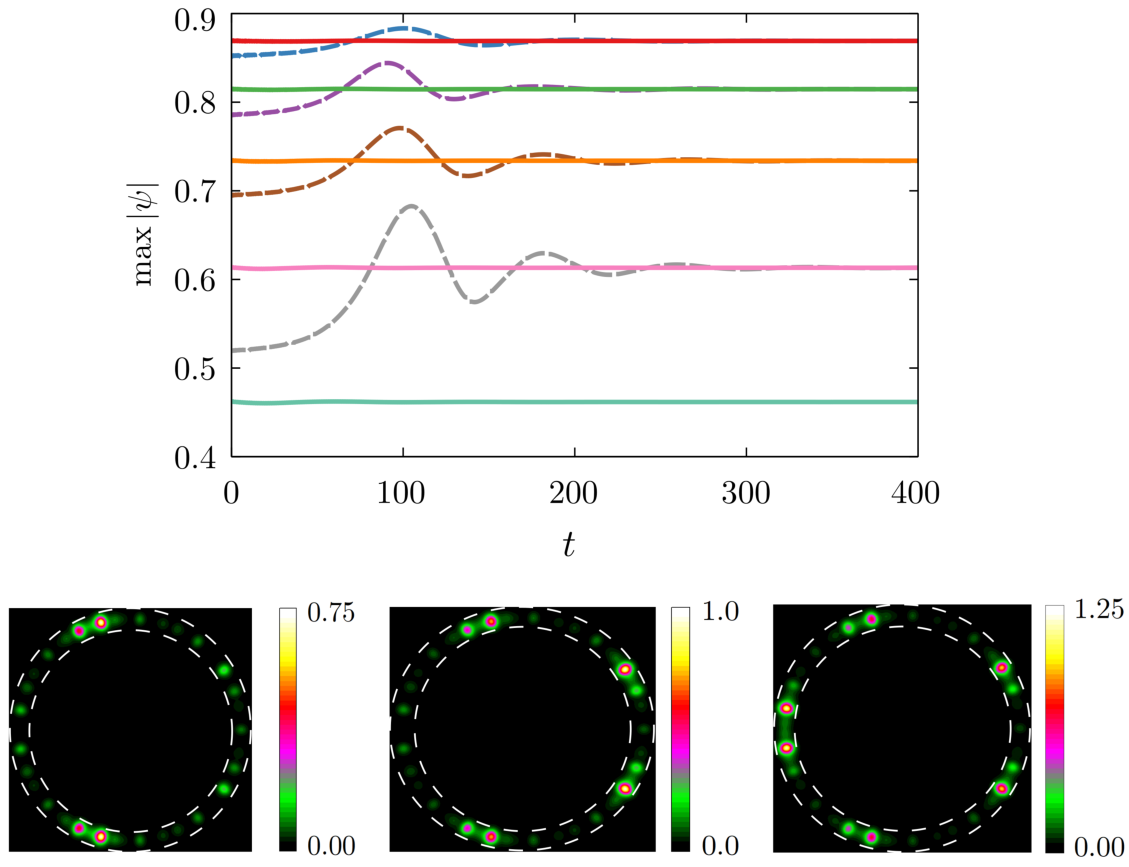}
 	\end{center}
 	\caption{Upper panel shows temporal dynamics of maximal amplitude $\max_{(x,y)} |\psi|$  for nine nonlinear states coexisting at the pump frequency $\varepsilon=-10.59$ for pump amplitude $h_0=0.006$ [see dotted vertical line in Fig.~\ref{fig:linresonances}(b)].  Solid and dashed curves correspond to the evolution of  stable and unstable states, respectively. Three lower panels show moduli $|\psi|$ for  for the three stable solutions with largest IPRs.}
 	\label{fig:dynamics}
 \end{figure}

In the linear system, each resonant peak is symmetric with respect to sufficiently small left and right detunings from the  frequency corresponding to the center of the peak, where the local maximum of the excited amplitude is achieved. Therefore  no more than one   linear state can be resonantly excited at each frequency. However, when the nonlinearity is taken into account, the resonant  peaks become distinctively asymmetric, as illustrated in  Fig.~\ref{fig:linresonances}(b) for two different pump amplitudes $h_0$, compare left and right vertical axes in this figure. Only a part of the picture corresponding to the resonances at relatively small frequencies is shown in  Fig.~\ref{fig:linresonances}(b). Since the degree of asymmetry of resonance peaks strengthens    with  the increase of  $h_0$, for a sufficiently large  pump amplitude   two or more nonlinear states   coexist at the same value of the pump frequency $\varepsilon$, which can result in bi- or multistability. To illustrate the stability properties of localized states, we present the  dynamical modelling results for nine nonlinear states coexisting at the pump strength $h_0=0.006$ and frequency   $\varepsilon=-10.59$ [marked with  the vertical dotted line in Fig.~\ref{fig:linresonances}(b)]. This figure indicates that there   are five multistable states at the chosen frequency which are always situated at upper   branches of tilted resonant peaks (see the evolutions corresponding to solid lines in Fig.~\ref{fig:dynamics}).  Other four states are unstable, and   each unstable mode   dynamically switches into the  stable  mode at   the corresponding upper branch.  Therefore stable states feature the attractor-like behavior.

\rev{The results plotted in the inset in Fig.~\ref{fig:linresonances}(b) indicate that for   stronger pump strengths the IPRs of the excited states tend to  decrease. This behavior can be explained by the nonlinearity-induced hybridization between     several localized states, which results in the excitation of patterns composed of several pairs of bright spots. Three examples of such hybridized states are shown in the bottom row of Fig.~\ref{fig:dynamics}. Solutions with larger maximal amplitude contain more bright peaks than those with small amplitudes.  This trend in general agrees with the repulsive character of the polariton interactions which are expected to promote the delocalization of  strongly nonlinear high-amplitude states.}

\begin{figure}%[h]
	\begin{center}		
 \includegraphics[width=1\columnwidth]{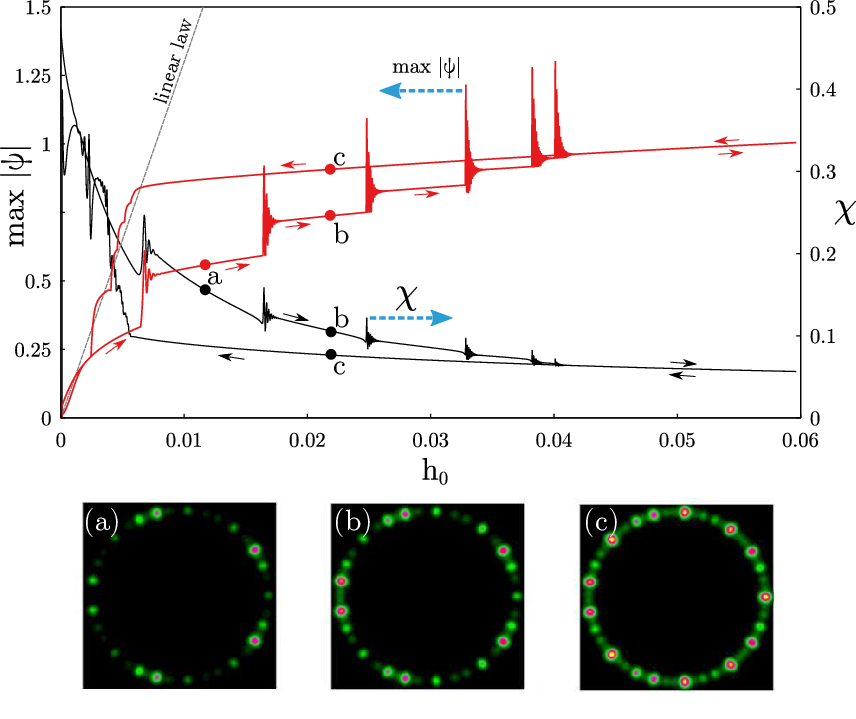}%
	\end{center}
	\caption{ \rev{The upper panel shows the evolution of the maximal amplitude (left vertical axis) and IPR  (right vertical axis) as the pump amplitude  changes in time (at first, increases from $h_0\approx0$ to $h_0\approx 0.06$ and   then decreases from $h_0\approx 0.06$  to  $h_0\approx0$). Small arrows correspond to the increase of time and highlight the hysteresis-like response. Points  (a,b,c) in the upper panel correspond to the solutions shown in the lower row. The results are computed  using the numerical integration of Eq.~(\ref{eq:gpe}) in  the nonlinear regime, at  the zero-charged  pump at frequency $\varepsilon = -10.647$. } } 
	\label{fig:hyst}
\end{figure}

\rev{The nonlinear response of the system can also be probed using the adiabatic increase of the pump intensity  as illustrated in Fig.~\ref{fig:hyst}.  In the regime of weak pump, the excited  amplitude   grows linearly with the pump strength. However, the further increase of the pump strength eventually leads to the deviation from the linear growth law and triggers a cascade of jumps from low-amplitude to high-amplitude  stable states   (in Fig.~\ref{fig:hyst}, these jumps are separated by spiky transients).  The increase of the pump amplitude $h_0$ is  accompanied by the general decay of the IPR $\chi$, which results from the excitation of additional bright spots in the high-amplitude solution [compare plots (a) and (b) in the lower row of Fig.~\ref{fig:hyst}]. The multistable nature of the system  manifests in a hysteresis-like behavior~\cite{Rosanov}:   when the pump amplitude is switched from increasing to decreasing, the system  undergoes a different evolution (see small arrows in Fig.~\ref{fig:hyst} that highlight  different paths followed by the system as the pump amplitude $h_0$ increases and decreases). The upper branch of the amplitude curve in Fig.~\ref{fig:hyst}  corresponds to the lower branch of the IPR curve.}

  \begin{figure}%[h]
	\begin{center}		
 \includegraphics[width=1\columnwidth]{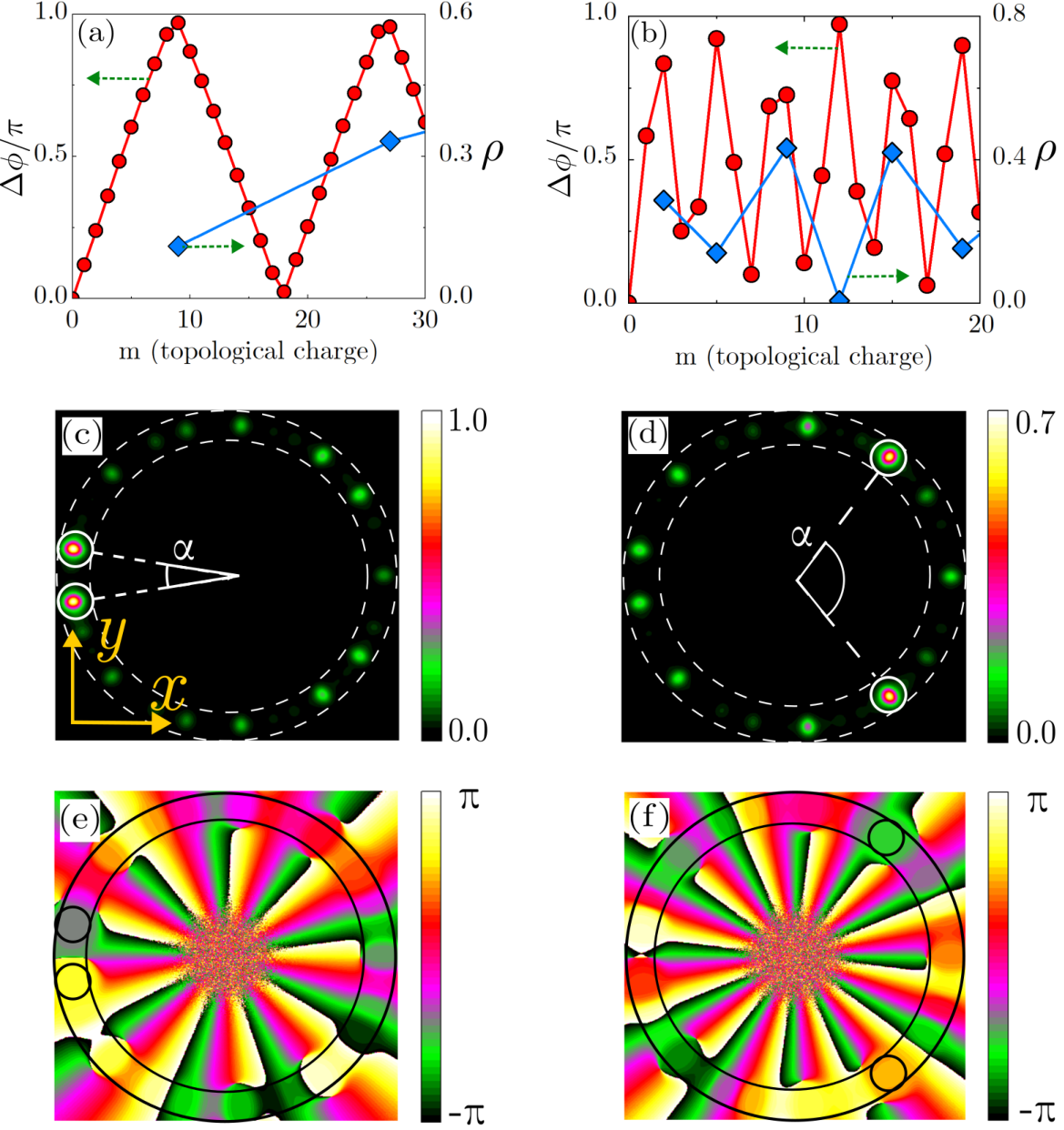}%
	\end{center}
	\caption{(a,b) Phase difference $\Delta\phi$ (in units of $\pi$) and roundoff distance $\rho$ as functions of topological charge $m$ for two different out-of-phase states with $\alpha\approx 20^\circ$ (a) and $\alpha\approx 105^\circ$ (b); panels (c,d) and (e,f)  show modulus and phase distributions of  out-of-phase states exited with vortical resonance pump with $m=9$ (c,e) and  $m=12 $ (d,f). This figure corresponds to the linear regime. %Here the pump beam amplitude $h_0= 0.01$.  
 }
 %eps = -10.889342773826197; eps = -10.819739586125417;
	\label{fig:phases}
\end{figure}

\section{Excitation of out-of-phase states with vortical pump}
\label{sec:out}

As  noted above in Sec.~\ref{sec:linear}, the spectrum of linear localized states includes multiple pairs of in-phase and out-of-phase two-peaked modes which are situated at relatively close energies. The zero-charged resonant pump (i.e. that with $m=0$) dictates that the phase of the exited pattern is uniform along the ring and therefore it cannot excite out-of-phase modes. However, the latter can be excited using a  vortical   pump with a properly chosen topological charge $m$. Let $\alpha \in (0, \pi)$ be the angle formed by the two radii connecting the center of the ring and the two out-of-phase spots in a given solution (see illustration in Fig.~\ref{fig:phases}). The phase difference between these peaks is equal to  to $\Delta\phi_1 = (2q+1)\pi$, where $q=0,1,\ldots$ is an unknown integer. At the same time, the pump beam with topological charge equal to $m$ induces the phase incursion  equal to   $\Delta \phi_2 = m\alpha$ over the angle $\alpha$. The ``optimal'' situation for the excitation of the purest out-of-phase state would correspond to the exact equality $\Delta\phi_1 = \Delta \phi_2$. However, since $m$ is integer, this equality, generically speaking, can be satisfied only approximately. To find a topological charge which corresponds to the purest out-of-phase mode, we introduce an auxiliary quantity $\mu_q = (2q+1)  \pi/\alpha$ and for each $q$ measure the distance between $\mu_q$ and the closest integer   by introducing $\rho_q = |\mu_q - [\mu_q]|$, where $[\mu]$ means  round towards nearest integer.  The optimal situation takes place if for some $q$ one has $\rho_q=0$. Then the out-of-phase state can be excited with the topological charge $m = \mu_q$.  To check this prediction, we perform the direct simulation of excited states for different topological charges (limiting the consideration by $m$ ranging from 0 to 30) and measure the phase difference $\Delta\phi$ between  the two peaks. In Fig.~\ref{fig:phases}(a,b) we juxtapose  the dependencies $ \Delta\phi /  \pi $ \textit{vs.} $m$ and $\rho_q$ \textit{vs.}  $[\mu_q]$ for two different out-of-phase states. These results confirm that the purest out-of-phase states with  $\Delta\phi\approx \pi$ are indeed excited by topological charges that correspond to small values of $\rho_q$. 
%This understanding allows one to control the phase distribution of     localized states. 
It should be emphasized that, since different out-of-phase states correspond to different angles $\alpha$, the  optimal topological charges are also generically different.
%, and for  out-of-phase states with  small angle $\alpha$  it is more difficult to find a topological charge that excites a nearly out-of-phase   state. 
%For there is only one topological charge $m=9$ that can be used (however, other possible topological charges can be potentially found for $m>20$ which are not considered in the figure)

\section{Conclusion}
\label{sec:concl}

We  have  presented a theoretical study for localization of exciton-polaritons in annular microcavities which have an additional quasiperiodic modulation implemented as a bicosine dependence. We have started with the localization diagram in the linear and conservative regime, demonstrated that the cavity can feature a sharp mobility edge, and classified the localized states. Then it has been shown that the localized patterns can be excited using  a resonant pump and feature distinctive nonlinear behavior  which results in multistability, i.e.,  in the coexistence    of  several dynamically stable states at the same frequency of resonant pumping. \rev{The complexity of the nonlinear resonances   diagram, i.e., the number of coexisting stable and unstable branches, can be controlled by changing the intensity of the incident pump beam.} The multistable states dynamically behave as attractors, and  unstable states dynamically switch to their stable counterparts from upper branches of tilted   resonance peaks. \rev{A nonmonotonic change of the pump amplitude results in a hysteresis, i.e., the nonequivalence of paths followed by the system as the pump amplitude increases and decreases.} Most of the localized states exist in the form of in-phase and out-of-phase dipoles, and the latter one can be excited by tuning the topological charge of the vortical pump.

We expect that our results will promote further research of nonlinear effects for exciton-localized polariton states that form in quasicrystalline and disordered media. One  relevant direction of future work is related to a thorough analysis of polarization   effects  that can be taken into account in the form of the TE-TM splitting. The pseudospin degree of freedom may considerably enrich the behavior of localized states. Its effect can be especially interesting if considered together with  an  external magnetic field applied to the microcavity. Another relevant issue is the excitation of localized state using an incoherent, off-resonant pumping scheme which is expected to enable more complex dynamical behavior, such as the nonlinear symmetry-breaking or dynamical oscillations between different stable localized modes, similar to the multimode dynamics recently explored for atomic Bose-Einstein condensates in one-dimensional bichromatic optical lattices \cite{Prates}. \rev{It would also be interesting to study the role of the repulsive polariton interaction in the context of a finite-temperature phase transition between fluid and insulator that was theoretically predicted for disordered weakly interacting bosons in Ref.~\cite{Aleiner}.}

\begin{acknowledgments}

The authors are grateful to Igor Chestnov and Alexey Yulin for valuable discussions. This work was supported by the Ministry of Science and Higher Education of Russian Federation (``goszadanie no. 2019-1246''), and by  ``Priority 2030 Academic Leadership Program''.  

\end{acknowledgments}

%\bmsection{Funding}  Ministry of Science and Higher Education of Russian Federation, goszadanie no. 2019-1246. 

%\bmsection{Disclosures} The authors declare no conflicts of interest.

%\bmsection{Data availability} No data were generated or analyzed in the presented research.

\end{document}